\documentclass[sigconf]{acmart}
\AtBeginDocument{%
  }

\setcopyright{acmlicensed}
\copyrightyear{2026}
\acmYear{2026}
\acmDOI{XXXXXXX.XXXXXXX}
\acmISBN{978-1-4503-XXXX-X/2018/06}

\usepackage{accsupp}
\usepackage{booktabs}
\usepackage{multirow}
\usepackage{pifont}
\newcommand{\cmark}{\ding{51}}
\newcommand{\xmark}{\ding{55}}

\usepackage{xspace}

\newcommand{\datasetname}{MemeChain\xspace}

\providecommand{\eg}{\emph{e.g.,} }
\newcommand\mypara[1]{\noindent \textbf{#1}}

\begin{document}

\title{MemeChain: A Multimodal Cross-Chain Dataset for Meme Coin Forensics and Risk Analysis}

% MemeChain: A Multimodal Cross-Chain Meme Coin Dataset
% temporary title

\author{Alberto Maria Mongardini}
\orcid{}
%\affiliation{%
%  \institution{Sapienza University of Rome}
%  \city{Rome}
%  \country{Italy}
%}
%\email{mongardini@di.uniroma1.it}
\affiliation{%
  \institution{Technical University of Denmark}
  \city{Copenhagen}
  \country{Denmark}
}
\email{among@dtu.dk}

\author{Alessandro Mei}
\affiliation{%
  \institution{Sapienza University of Rome}
  \city{Rome}
  \country{Italy}
}
\email{mei@di.uniroma1.it}

%\renewcommand{\shortauthors}{Mongardini et al.}

\begin{abstract}
  The meme coin ecosystem has grown into one of the most active yet least observable segments of the cryptocurrency market, characterized by extreme churn, minimal project commitment, and widespread fraudulent behavior. While countless meme coins are deployed across multiple blockchains, they rely heavily on off-chain web and social infrastructure to signal legitimacy. These very signals are largely absent from existing datasets, which are often limited to single-chain data or lack the multimodal artifacts required for comprehensive risk modeling.

  To address this gap, we introduce \datasetname, a large-scale, open-source, cross-chain dataset comprising 34,988 meme coins across Ethereum, BNB Smart Chain, Solana, and Base. \datasetname integrates on-chain data with off-chain artifacts, including website HTML source code, token logos, and linked social media accounts, enabling multimodal and forensic study of meme coin projects. Analysis of the dataset shows that visual branding is frequently omitted in low-effort deployments, and many projects lack a functional website.
  Moreover, we quantify the ecosystem's extreme volatility, identifying 1,801 tokens (5.15\%) that cease all trading activity within just 24 hours of launch.
  By providing unified cross-chain coverage and rich off-chain context, \datasetname serves as a foundational resource for research in financial forensics, multimodal anomaly detection, and automated scam prevention in the meme coin ecosystem.
  
\end{abstract}





\maketitle

\section{Introduction}

Meme coins represent one of the most impactful phenomena in the cryptocurrency landscape. Originating as a satirical experiment with Dogecoin in 2013~\cite{nani2022doge}, the sector boasted a market capitalization exceeding \$117 billion by 2024~\cite{meme_success_cmc}. The primary allure of these assets lies in their potential for astonishing returns; prominent examples such as Pepe (400,000\%) and Shiba Inu (80,000,000\%)~\cite{impressive_return_meme_coins_spa} have cemented the narrative of overnight wealth. This sentiment, coupled with the emergence of low-barrier deployment platforms like \textit{pump.fun}~\cite{pump_success}, has triggered an unprecedented proliferation of new tokens.
Recognizing the importance of the phenomenon, regulatory bodies have moved to formalize these tokens. The U.S. Securities and Exchange Commission (SEC) now categorizes meme coins as "crypto assets inspired by internet memes, characters, current events, or trends," specifically noting that their value is derived from the promoter's ability to "attract an enthusiastic online community"~\cite{meme_coins_sec}. However, this definition highlights a critical systemic vulnerability: Unlike utility tokens anchored in technological function, meme coins rely almost exclusively on perceived social momentum to sustain value~\cite{nani2022doge}. This dependence creates a fertile ground for market manipulation, where artificial growth strategies are essential precursors to profit extraction~\cite{mongardini2025midsummer}.
The academic understanding of the meme coin lifecycle remains limited. Existing studies largely focus on financial time-series analysis or on-chain transaction graphs, often overlooking the infrastructure behind, specifically the web and social presence, that promoters use to legitimize their projects. A dataset that bridges the gap between on-chain mechanisms and off-chain marketing is notably absent.

To address this gap, we present \datasetname, a large-scale, cross-chain repository designed to facilitate forensic analysis of the meme coin ecosystem. 
To the best of our knowledge, \datasetname is the largest curated collection of meme coins to date, including over 34,988 meme coins across major blockchains such as Ethereum, BNB Smart Chain (BSC), Solana, and Base.
Moreover, \datasetname captures the lifecycle of meme coin projects, combining granular transaction timestamps with a unique archive of raw HTML source code from project websites and token logo images.

Our analysis of this data reveals a stark reality beneath the hype. We identify the pervasive phenomenon of One-Day Meme Coins, assets that cease all transactional activity within 24 hours of deployment. Furthermore, we find that web infrastructure serves as a potent proxy for project viability. A significant portion of projects launch with no web presence at all, and among those that do, we observe a pattern where websites are rapidly abandoned or rendered unreachable shortly after the initial liquidity event. By providing this cross-sectional snapshot of the ecosystem, \datasetname offers researchers the necessary ground truth to train early-warning systems and further investigate the ecosystem.

This paper makes the following contributions:

\begin{itemize} 
    \item \textbf{A Large-Scale Multi-Modal Dataset.} 
    We release the first open-source dataset linking on-chain data with off-chain web artifacts. Totaling over 1.46 GB, the collection comprises core metadata for 34,988 meme coins, enriched with a repository of token logos and website source code for a substantial subset of projects, enabling new forms of multimodal analysis.

    \item \textbf{Lifecycle Analysis.} We provide a granular analysis of token longevity, identifying the rapid abandonment of many projects and introducing the concept of One-Day Meme Coins, assets that lose all transactional volume within 24 hours.

    \item \textbf{Web Infrastructure Forensics.} We conduct the first systematic study of meme coin web presence, revealing that a significant proportion of projects lack a functional website or rely on ephemeral, low-cost hosting solutions. %We map the "digital silence" of abandoned domains, showing how web stability serves as a critical proxy for project legitimacy.

    \item \textbf{Roadmap for Future Research.} We outline how the dataset enables the training of survival analysis models, the detection of early-stage scam indicators, and the development of automated risk assessment tools.
\end{itemize}

%\section{Background}
%\input{sections/background}

\section{Related Work}
\mypara{Meme Coins: Culture and Economics.}
Internet memes, defined by Shifman~\cite{shifman2013memes} as ``units of popular culture that are circulated, imitated, and transformed by individual Internet users,'' have evolved into significant cultural phenomena. The intersection of this culture with blockchain technology gave rise to meme coins, which differ fundamentally from traditional cryptocurrencies. While standard assets often rely on technical utility, meme coins are characterized by their community-driven nature, humor incorporation, and viral origins~\cite{stencel2023meme, nani2022doge}.
Recent studies have begun to analyze the unique economic drivers of this sector. Kim et al.~\cite{kim2024identifying} demonstrated how social media sentiment directly drives meme coin valuation, while Belcastro et al.~\cite{belcastro2023enhancing} highlighted that meme coin price movements exhibit distinct volatility patterns requiring specialized prediction approaches. Unlike established assets (\eg Bitcoin), the value of meme coins is often decoupled from technological development, making them a unique asset class for behavioral finance study.

\mypara{Market Manipulation.}
The lack of regulatory oversight in the cryptocurrency ecosystem has fostered various forms of market manipulation~\cite{gandal2018price, krafft2018experimental, cernera2023ready, daian2020flash, cernera2024blockchain}.
Wash trading, a practice used to generate artificial volume, has been widely documented in ERC-20 tokens~\cite{victor2021detecting} and non-fungible token (NFT) marketplaces~\cite{von2022nft, la2023game}.
Pump and dumps are among the most prevalent manipulations and involve inflating an asset's price through misleading statements and coordinated buying, followed by selling the overvalued assets to unsuspecting investors~\cite{kamps2018moon, xu2019anatomy}. Studies examining the coordination of such schemes~\cite{hamrick2019economics,la2023doge,la2020pump} identified Telegram as the primary hub, a finding supported by platform-specific research~\cite{la2023sa,la2018pretending,la2025tgdataset}.
Finally, rug pulls function as exit scams where developers abruptly abandon projects, draining liquidity~\cite{zhou2024stop}. Recent work has highlighted the speed of these scams, specifically "1-day rug pulls," which have generated over \$240 million in illicit profits on BSC alone~\cite{cernera2023ready}.

\mypara{Meme Coin Datasets.}
While the mechanics of manipulation are well-studied, curated data specifically for meme coins remains scarce. Most existing datasets focus on Ethereum or specific scam types without distinguishing the meme coin sub-sector. 
For example, Wu et al.~\cite{wu2024tokenscout} have recently introduced TokenScout, a large-scale dataset (214k tokens) focusing on detecting rug pulls and honeypots within the Ethereum ecosystem. 
However, research specifically targeting meme coins is nascent. Long et al. recently introduced Coin-Meme~\cite{long2024bridging} and CoinVibe~\cite{long2024coinclip}, multimodal datasets sourced from the \textit{pump.fun} platform on Solana. These works leverage visual data (logos), textual narratives, and community sentiment to assess token viability and cultural alignment.

\mypara{Bridging the Gap.}
Despite these advances, current literature suffers from three critical limitations. First, existing datasets often focus on viability or successful tokens. By contrast, \datasetname explicitly captures also ephemeral assets, including One-Day Meme Coins and failed projects, providing a more realistic representation of the ecosystem's high churn and risk. Second, previous meme coin datasets are largely limited to Solana or Ethereum. Our work provides the first unified cross-chain view (Ethereum, BSC, Base, and Solana), enabling researchers to study migration patterns and comparative scam prevalence across different architectures. Finally, with 34,988 tokens, \datasetname is nearly an order of magnitude larger than previous multimodal collections (3,751 in~\cite{long2024bridging} and 6,231 in~\cite{long2024coinclip}), providing the data needed to train robust models.

\section{Data Collection}
\begin{figure}[!htbp]
    \centerline{\includegraphics[width=0.46\textwidth]{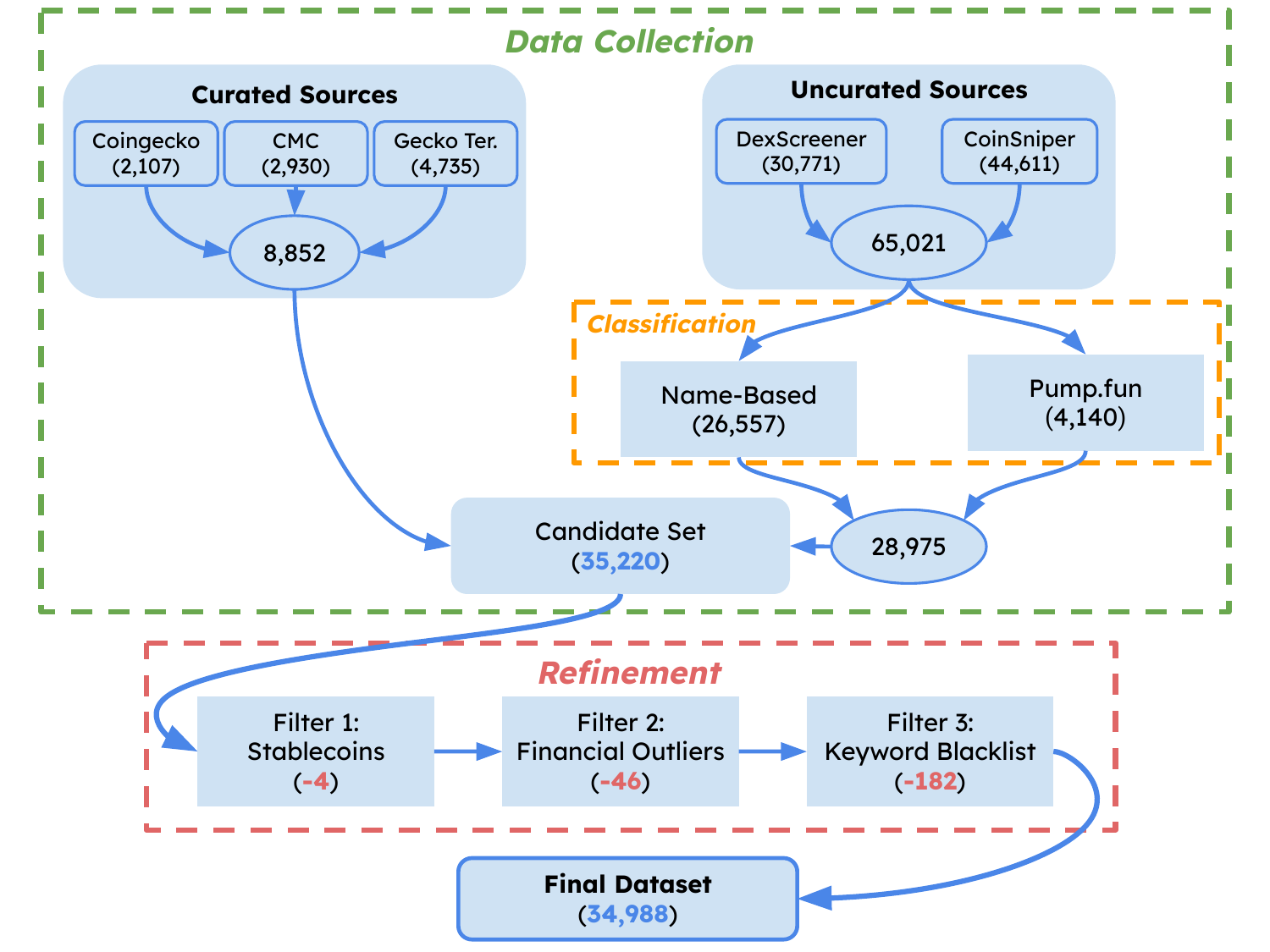}}
    \caption{The Data Collection Pipeline. Negative numbers indicate the count of tokens filtered out during the refinement.}
    \Description{A plot showing the data collection pipeline.}
  \label{fig:data_pipeline}
\end{figure}

To build the \datasetname dataset, we developed a multi-stage data collection pipeline (illustrated in Fig.~\ref{fig:data_pipeline}) to capture the full spectrum of the meme coin ecosystem, from established high-market-cap assets to newly deployed tokens. 

\subsection{Data Collection}
We employed a hybrid sourcing approach that combines verified data from curated cryptocurrency aggregators with comprehensive monitoring of decentralized exchange (DEX) platforms. This blockchain-agnostic strategy enables us to track cross-chain migration patterns and observe how meme coin creators leverage different blockchain ecosystems, insights that would be missed in single-chain analyses.

\subsubsection{Established Meme Coins}
\label{subsubsec:know_meme_coins}
Our initial approach focused on established cryptocurrency aggregators that explicitly categorize tokens as meme coins, providing a foundation of verified assets. We collected data from three primary sources.

\mypara{CoinMarketCap~\cite{coinmakretcap}} and \textbf{CoinGecko~\cite{coingecko}.}
These leading cryptocurrency tracking platforms maintain dedicated meme token categories. CoinGecko defines meme coins as "tokens with no intrinsic value, but with a strong community that fosters social interaction,"~\cite{meme_coins_definition_coingecko} while CoinMarketCap describes them as cryptocurrencies that "originated from internet memes...often created as a joke or for the purpose of satire...with community-driven nature and humorous themes~\cite{meme_coins_definition_coinmarketcap}." Both definitions align with the SEC's characterization~\cite{meme_coins_sec}, which we adopt as our formal definition. Using their APIs~\cite{coinmakretcap_api,coingecko_api}, we collected 2,930 tokens from CoinMarketCap and 2,107 from CoinGecko.

\mypara{Gecko Terminal~\cite{geckoterminal}.} 
This platform categorizes tokens by meme trends, reflecting the definition of meme coins as assets inspired by Internet culture~\cite{meme_coins_sec}. We developed a custom scraper to collect 4,735 tokens from pools tagged with relevant meme categories (animal, dog, cat, inu, and pump.fun).

After removing duplicates across these sources, we obtained 8,852 distinct verified meme coins.

\subsubsection{Emerging Tokens}
\label{subsubsec:new_dex_meme_coins}
To capture emerging tokens, we developed scrapers for two DEX aggregators: DexScreener~\cite{dexscreener} and CoinSniper~\cite{coinsniper}. These platforms employ more permissive listing criteria than established aggregators~\cite{coingecko_listing, coinmarketcap_listing}. DexScreener automatically indexes tokens upon liquidity pool creation and initial transaction activity~\cite{dexscreener_listing}, while CoinSniper requires only a basic application for immediate listing~\cite{coinsniper_listing}. We collected 30,771 tokens from DexScreener and 44,611 from CoinSniper, yielding 65,021 unique tokens.
Since DEX aggregators could list various token types beyond meme coins, we needed classification criteria to distinguish meme coins from conventional cryptocurrencies and utility tokens.

\paragraph{Classification Methodology}
To specifically identify meme coins, we adhere to the formal definition provided by the SEC, which characterizes them as "crypto assets inspired by internet memes, characters, current events, or trends"~\cite{meme_coins_sec}.
In particular, we observed that these cultural and thematic influences are consistently reflected in the token names, which exhibit distinctive linguistic patterns. These patterns can be systematically identified, and based on this insight, we developed a name-based classification method to detect meme coins within a larger dataset.

\begin{figure}[!htbp]
    \centerline{\includegraphics[width=0.46\textwidth]{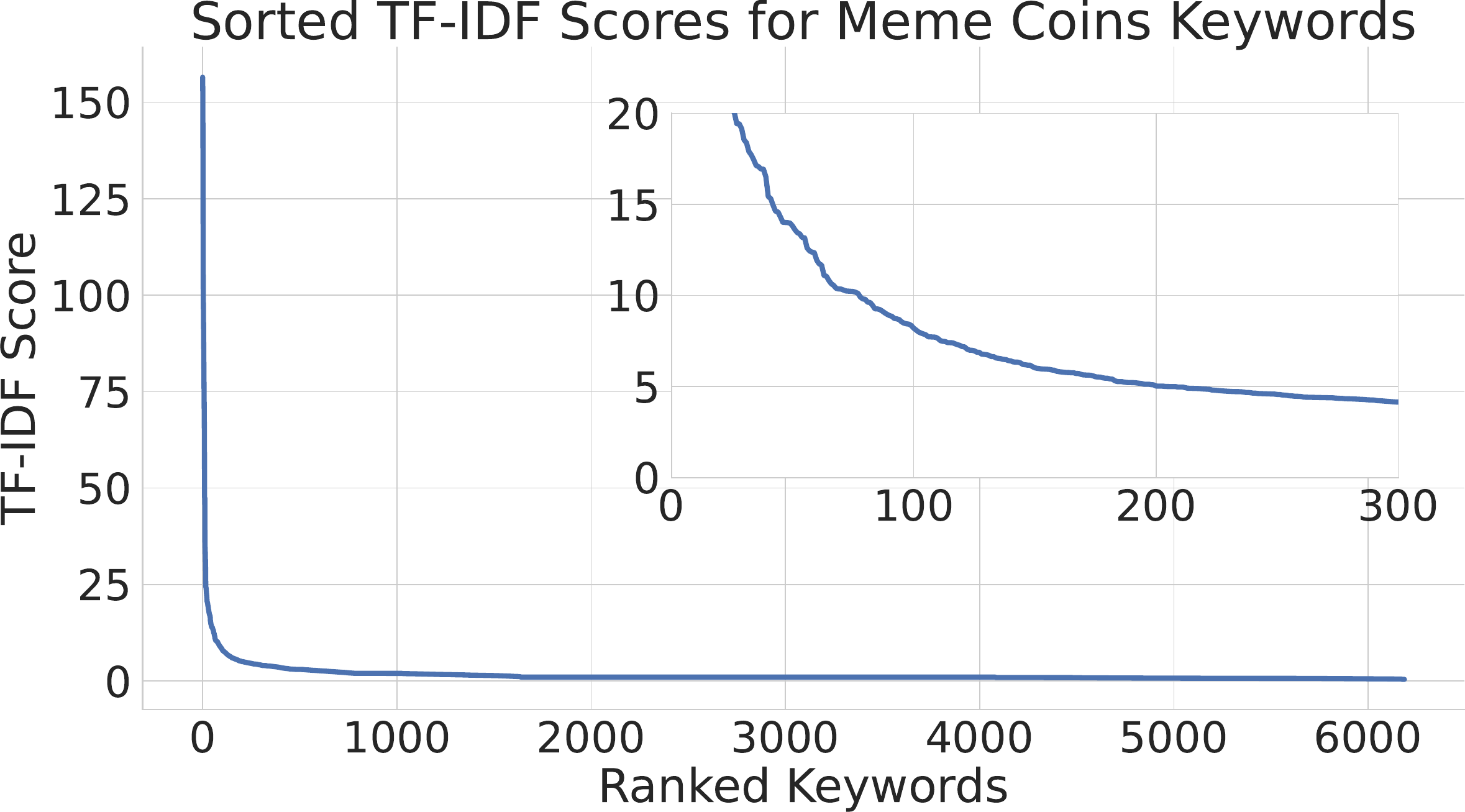}}
    \caption{TF-IDF scores of keywords associated with meme coin names.}
    \Description{A plot showing the TF-IDF scores of keywords extracted from meme coins' names.}
  \label{fig:tf_idf_names}
\end{figure}

\mypara{Name-Based Classification.}
We begin by examining the names of verified meme coins, as detailed in Sec.~\ref{subsubsec:know_meme_coins}, to identify common keywords associated with meme tokens. To extract these keywords, we apply Term Frequency–Inverse Document Frequency (TF-IDF) analysis to our dataset of verified meme tokens. This technique assigns higher weights to terms that frequently appear in meme coin names but are relatively rare in general cryptocurrency naming conventions~\cite{ramos2003using}.
Our methodology is as follows: First, we preprocess the token names by converting them to lowercase and removing special characters. Then, we tokenize the names into individual words (unigrams) and compute their TF-IDF scores. To determine the optimal number of keywords to retain, we use the elbow method, which helps identify the point at which additional keywords provide diminishing returns in terms of value~\cite{bholowalia2014ebk}. As shown in Fig.~\ref{fig:tf_idf_names} (Appendix), the elbow method suggests 150 keywords as the optimal cutoff. We then manually review these candidates, removing generic cryptocurrency terms (\eg SOL or ETH) that appear due to cross-chain versions of the same token (\eg Dogecoin (SOL)). This refinement results in a set of 126 meme-specific keywords.
The final keyword set reflects the thematic dominance within the meme coin market: animal-related terms (\eg \textit{dog}: 7.24\%, \textit{cat}: 6.92\%, \textit{inu}: 4.82\%) and trend-based terms (\eg \textit{AI}: 4.96\%) are heavily represented.
We apply this name-based classification to the dataset collected in Sec.~\ref{subsubsec:new_dex_meme_coins}, identifying tokens as meme coins if their names contain any of the 126 meme-specific keywords. Using this method, we detect 26,557 meme coins.

\mypara{Pump.fun-Based Classification.}
In addition to the name-based classification, we directly include tokens from the \textit{pump.fun} platform, identifiable by the \textit{.pump} suffix in their addresses, for two reasons. First, \textit{pump.fun} explicitly markets itself as a platform dedicated to meme coin creation~\cite{pump_fun_platform}, suggesting a high probability that tokens listed there are meme-related. Second, to empirically validate this, we manually inspected 100 randomly sampled tokens from the \textit{pump.fun} platform. Our inspection confirmed that 100\% of these tokens featured names referencing internet memes, pop culture, or viral trends, which aligns with the SEC's definition~\cite{meme_coins_sec}. This manual review also revealed niche memes that were not captured by the name-based classification, helping to mitigate false negatives and ensuring emerging trends are included.
Using this approach, we identify an additional 4,140 meme coins, bringing the total to 28,975 meme coins.

By combining the known meme coins from Sec.~\ref{subsubsec:know_meme_coins} with the newly identified tokens, we obtain a total of 35,220 meme coins.

\subsubsection{Dataset Refinement}
\label{subsubsec:dataset_refinement}
Manual inspection of tokens with high prices or large market capitalizations reveals a small number of false positives, such as stablecoins and wrapped assets. To address this issue, we apply a three-stage filtering procedure to remove misclassified tokens and retain only meme coins. Our refinement strategy is intentionally conservative: We prioritize precision over recall, accepting the exclusion of some meme coins to minimize false positives and improve the robustness of subsequent analyses.

First, we remove stablecoins using CoinMarketCap’s API~\cite{coinmakretcap_api}. By leveraging CoinMarketCap’s list of 82 labeled stablecoins, we identify and discard 4 tokens, reducing the dataset to 35,116 tokens.

\begin{figure}[!htbp]
    \centerline{\includegraphics[width=0.46\textwidth]{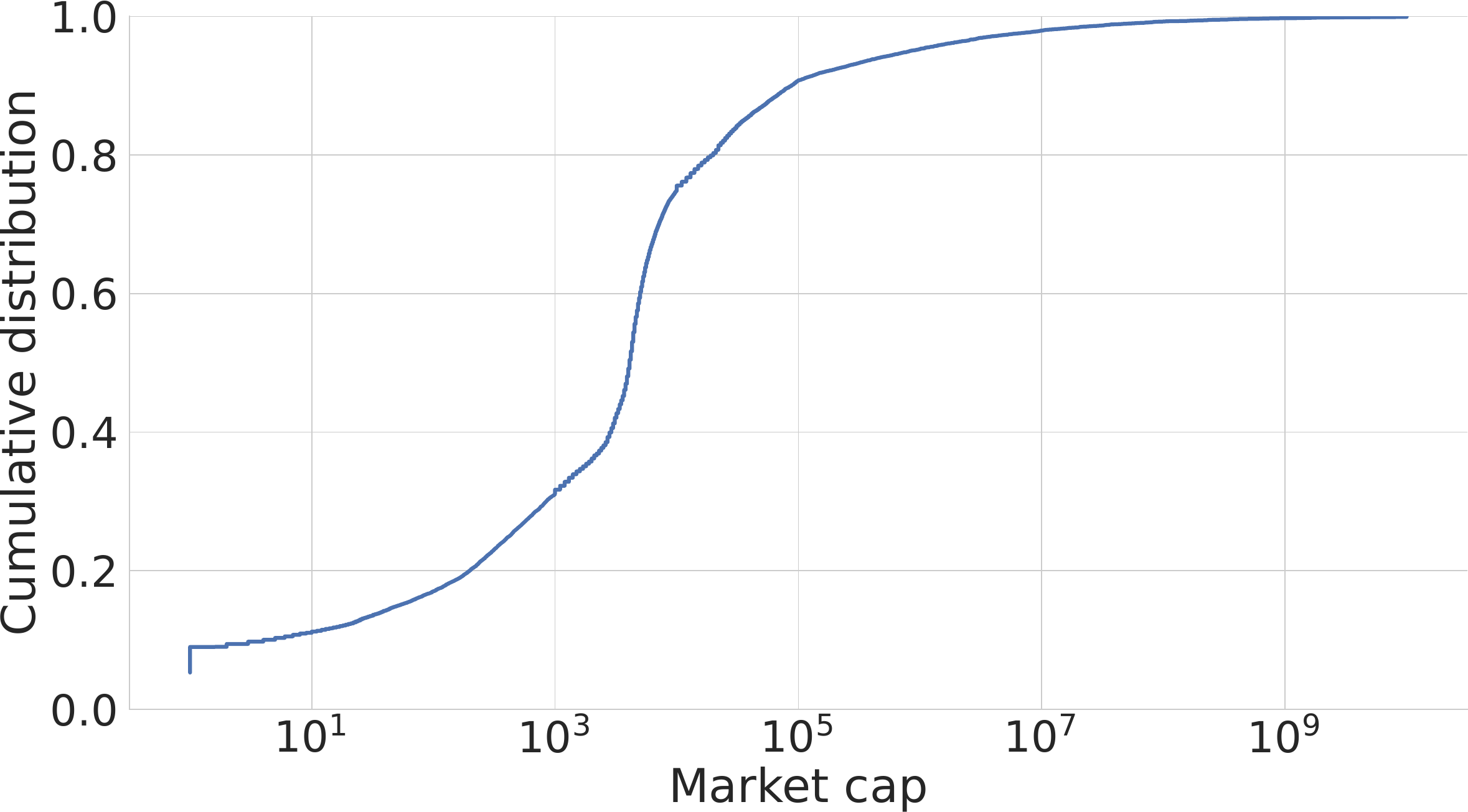}}
    \caption{CDF of the market cap. of meme coins collected.}
    \Description{A plot showing the CDF of the market capitalization of meme coins collected.}
  \label{fig:cdf_dataset_refinement}
\end{figure}

In the second stage, we target non-meme tokens with substantial market presence. Rather than applying generic outlier detection based solely on price or market capitalization, which could incorrectly exclude high-value meme coins, we adopt a targeted filtering strategy. Specifically, we flag tokens that satisfy at least one of the following conditions: (i) a price exceeding \$0.80, as meme coins typically trade at low prices~\cite{nani2022doge}, and this threshold helps capture stablecoins that may be slightly depegged; or (ii) a market capitalization above $10^7$, since the cumulative distribution function (CDF) shown in Fig.~\ref{fig:cdf_dataset_refinement} indicates that most tokens in our dataset fall below this level.
This step identifies 525 tokens for further inspection. We then query CoinGecko’s API~\cite{coingecko_api} to retrieve their associated categories and remove tokens labeled with non-meme categories (\eg Staking, Bridged). As a result, 46 additional tokens are excluded.

Finally, we apply a conservative string-matching filter to remove tokens whose names contain terms associated with non-meme assets, specifically \textit{usd}, \textit{wrapped}, and \textit{staked}. This step identifies and excludes 182 tokens, yielding a final dataset of 34,988 meme coins, which we publicly release~\cite{dataset_repository}. Fig.~\ref{fig:data_pipeline} summarizes the collected tokens and the successive filtering steps used to refine the dataset.

\subsection{Dataset Overview}
\label{subsec:dataset_overview}
Our data collection was conducted during mid-October 2024, capturing a comprehensive snapshot of the meme coin ecosystem.

\subsubsection{Blockchain Distribution}
The analysis of the 34,988 collected meme coins reveals clear platform-level preferences. BNB Smart Chain (BSC) hosts the largest share, with 16,405 tokens (46.88\%), followed by Solana with 11,829 tokens (33.8\%), Ethereum with 3,814 tokens (10.9\%), and Base with 828 tokens (2.36\%).

Ethereum~\cite{ethereum_doc} pioneered programmable smart contracts, establishing the foundation for decentralized applications. However, its high transaction fees has motivated the emergence of alternative architectures. One response has been the development of new Layer~1 blockchains such as BNB Smart Chain (BSC)~\cite{bnbsmartchain}, which is EVM-compatible and offers lower transaction costs. Another approach is the deployment of Layer~2 solutions built on Ethereum itself, such as Base~\cite{base_doc}, which improves scalability while remaining within the Ethereum ecosystem. In contrast, Solana~\cite{solana_doc} adopts a non-EVM architecture, emphasizing low fees and high throughput through its proof-of-history consensus mechanism.

These four blockchains account for 32,876 meme coins, representing 93.96\% of the dataset. Given their dominance, all subsequent analyses focus exclusively on tokens deployed on these platforms.

\subsubsection{Addresses Validation}
\label{sec:address_validation}
Next, we validate token addresses to remove invalid entries and non-deployed contracts, such as placeholders (e.g., \textit{0x000comingsoon} or \textit{upcoming}).
For EVM-compatible blockchains (BSC, Ethereum, and Base), we first perform format validation: addresses must be 42 characters long, begin with ``0x'', and contain only hexadecimal characters. This step detects 791 invalid addresses (739 on BSC, 51 on Ethereum, and 1 on Base). For addresses that pass this check, we query blockchain explorer APIs (Etherscan~\cite{etherscan_api} for Ethereum, BscScan~\cite{bscscan_api} for BSC, and BaseScan~\cite{base_scan} for Base) to verify the existence of contract-creation transactions. This process identifies 254 addresses (211 on BSC and 43 on Ethereum) that lack deployment transactions and are therefore not valid contracts.
For Solana tokens, we validate mint addresses using Solana’s RPC API~\cite{solana_rpc}, identifying 20 invalid entries.

Tab.~\ref{tab:validation_results} summarizes the validation procedure. After removing a total of 1,065 invalid or non-deployed addresses, the final dataset comprises 31,811 valid tokens (96.76\% of the original set): 15,455 on BSC, 11,809 on Solana, 3,720 on Ethereum, and 827 on Base.

\begin{table}[h]
\centering
\small
\caption{Summary of token address validation process}
\label{tab:validation_results}
\begin{tabular}{lcccc}
\toprule
\textbf{Platform} & \textbf{Invalid} & \textbf{Undeployed} & \textbf{Removed} & \textbf{Retained} \\
\midrule
BSC       & 739  & 211  & 950  & 15,455 \\
Ethereum  & 51   & 43   & 94   & 3,720  \\
Solana    & -    & 20   & 20   & 11,809 \\
Base      & 1    & 0    & 1    & 827    \\
\midrule
\textbf{Total} & \textbf{791} & \textbf{254} & \textbf{1,065} & \textbf{31,811} \\
\bottomrule
\end{tabular}
\end{table}

\subsubsection{Dataset Schema}
Tab.~\ref{tab:dataset_schema} details the schema of \datasetname, categorizing features into three logical groups: technical identifiers, social presence, and market metrics.

\mypara{Technical Specifications.} 
The technical data includes the token name, symbol, image, creation date, contract address, and blockchain. 
The core identifiers (contract address, blockchain) facilitate cross-platform tracking. 
To support longitudinal analysis, we retrieved deployment timestamps for validated contracts using blockchain explorer APIs (Etherscan, BscScan, BaseScan) for EVM blockchains and Alchemy~\cite{alchemy_api} for Solana. 
Regarding visual assets, we successfully retrieved the logos for 15,095 tokens (43.14\%) using the same aggregators leveraged in the data collection phase. Since ERC-20 and SPL standards do not enforce on-chain image storage to maintain efficiency, logo management is an off-chain responsibility of the creator. Consequently, we deliberately retain tokens with missing logos, treating this absence not as missing data, but as a predictive binary feature potentially correlated with low-effort deployments and scam vectors.
%The remaining tokens could lack a corresponding icon. This distinction arises from the technical architecture of the ERC-20 and SPL token standards, which deliberately keep contracts lightweight for efficiency: Storing images on-chain would be prohibitively expensive. Consequently, token metadata is an off-chain concern that requires active management by token creators, which explains why newly created or less established tokens often appear without logos on various platforms. We deliberately retain these logo-less tokens in \datasetname. We posit that the presence or absence of a logo is not merely missing data, but a potentially predictive binary feature. We hypothesize these tokens may correlate with low-effort deployments, providing a valuable variable for future researchers to investigate low-commitment scam vectors.

\mypara{Social Presence.} 
This category captures the project's digital footprint. It includes the official website URL and handles for major communication channels (Telegram, Discord, and X). Moreover, we archived the raw HTML source code for all accessible websites.

\mypara{Market Metrics.} 
We collected economic information (price and market cap in USD) for 20,692 tokens (65.05\%). The remaining 11,119 tokens lack economic data. Specifically, these tokens were collected from DexScreener, CoinSniper, and Gecko Terminal, platforms that derive pricing information from liquidity pools where tokens are actively traded. Without an active liquidity pool at the time of collection, price and market capitalization cannot be determined, explaining the absence of economic data for these tokens.

\begin{table}[ht]
    \caption{Dataset Schema and Field Descriptions.}
    \label{tab:dataset_schema}
    \centering
    \small % Slightly smaller font to fit column width
    \begin{tabular}{c c c p{3.2cm}}
        \toprule
        \textbf{Category} & \textbf{Field Name} & \textbf{Type} & \textbf{Description} \\
        \midrule
        \multirow{5}{*}{\textbf{Technical}} 
          & \texttt{token\_address} & String & Unique contract address. \\
          & \texttt{chain} & String & Network.  \\
          & \texttt{token\_name} & String & Full token name. \\
          & \texttt{symbol} & String & Ticker symbol. \\
          & \texttt{image} & String & Path to the token image. \\
          & \texttt{deploy\_date} & Time & Contract creation date. \\
        \midrule
        \multirow{3}{*}{\textbf{Social}} 
          & \texttt{website\_url} & String & Official project URL. \\
          & \texttt{html\_path} & String & Path to HTML snapshot. \\
          & \texttt{X\_url} & String & X account. \\
          & \texttt{telegram\_url} & String & Telegram channel. \\
          & \texttt{discord\_url} & String &Discord server. \\
        \midrule
        \multirow{2}{*}{\textbf{Metrics}} 
          & \texttt{price\_usd} & Float & Price at collection time. \\
          & \texttt{market\_cap} & Float & Market capitalization.  \\
        \bottomrule
    \end{tabular}
\end{table}

\subsubsection{Comparison with Other Datasets}
\label{subsubsec:comparison}
To situate \datasetname within the broader landscape of cryptocurrency research, we compare it against datasets focusing on both meme coins and general token analysis. As detailed in Tab.~\ref{tab:comparison}, our dataset uniquely combines cross-chain coverage with granular off-chain artifacts.

\mypara{Meme Coin Datasets.}
Recent works like Coin-Meme~\cite{long2024bridging} and CoinVibe~\cite{long2024coinclip} have established the importance of multimodal analysis by collecting token logos and textual narratives from the Solana ecosystem. \datasetname matches this visual capability by providing token logos for our 34,988 assets while expanding the scope of off-chain data. While~\cite{long2024bridging, long2024coinclip} focus on platform-internal metrics (\eg Pump.fun comments), \datasetname captures the project's external web presence. We provide the archived HTML source code of the official project websites and mapped social media handles (X, Telegram, Discord). This allows researchers to analyze the landing page of a scam, where fraudulent claims and phishing mechanisms are often hosted, offering a dimension of fraud detection that logo analysis alone cannot provide.

\mypara{General Scam Datasets.}
Large-scale scam datasets like TokenScout~\cite{wu2024tokenscout} offer impressive volume (214k tokens) but lack domain specificity. They treat scams as generic anomalies on Ethereum, missing the nuanced social signaling that defines meme coin fraud. By integrating logos, website HTML, and social pages across four blockchains, \datasetname offers a more holistic view of the scam lifecycle than purely transaction-based datasets.

\begin{table}[ht]
    \centering
    \caption{Comparison of \datasetname with existing cryptocurrency datasets. %While matching the visual capabilities (Logos) of niche meme datasets, we uniquely provide the external web artifacts (HTML) and cross-chain breadth necessary for scam research.
    }
    \label{tab:comparison}
    \small
    \begin{tabular}{l c c c c c}
        \toprule
        \textbf{Dataset} & \textbf{Chain} & \textbf{Count} & \textbf{Logo} & \textbf{Web HTML} & \textbf{Social} \\
        \midrule
        TokenScout~\cite{wu2024tokenscout} & ETH & 214k & \xmark & \xmark & \xmark \\
        \midrule
        Coin-Meme~\cite{long2024bridging} & SOL & 3.7k & \cmark & \xmark & \cmark \\
        CoinVibe~\cite{long2024coinclip} & SOL & 6.2k & \cmark & \xmark & \cmark\\
        \midrule
        \textbf{\datasetname} & \textbf{Multi} & \textbf{35k} & \textbf{\cmark} & \textbf{\cmark} & \textbf{\cmark} \\
        \bottomrule
    \end{tabular}
    %\flushleft{\scriptsize{\textbf{Note:} Multi (4) = Ethereum, BSC, Base, Solana. "Web HTML" indicates the inclusion of archived source code from project websites.}}
\end{table}

\section{Dataset Analysis}

\subsection{Temporal Aspects}

\subsubsection{Temporal Evolution of Meme Coin Creation. }
\label{sec:temporal_evolution_of_memecoins}

\begin{figure}
    \centerline{%
    \Description{A detailed plot illustrating the creation dates of meme coins over time, showing temporal trends. It also displays their distribution across different blockchain platforms like Ethereum, Solana, and Binance Smart Chain, indicating which platforms host newer or more numerous meme coins.}
    \includegraphics[width=0.49\textwidth]{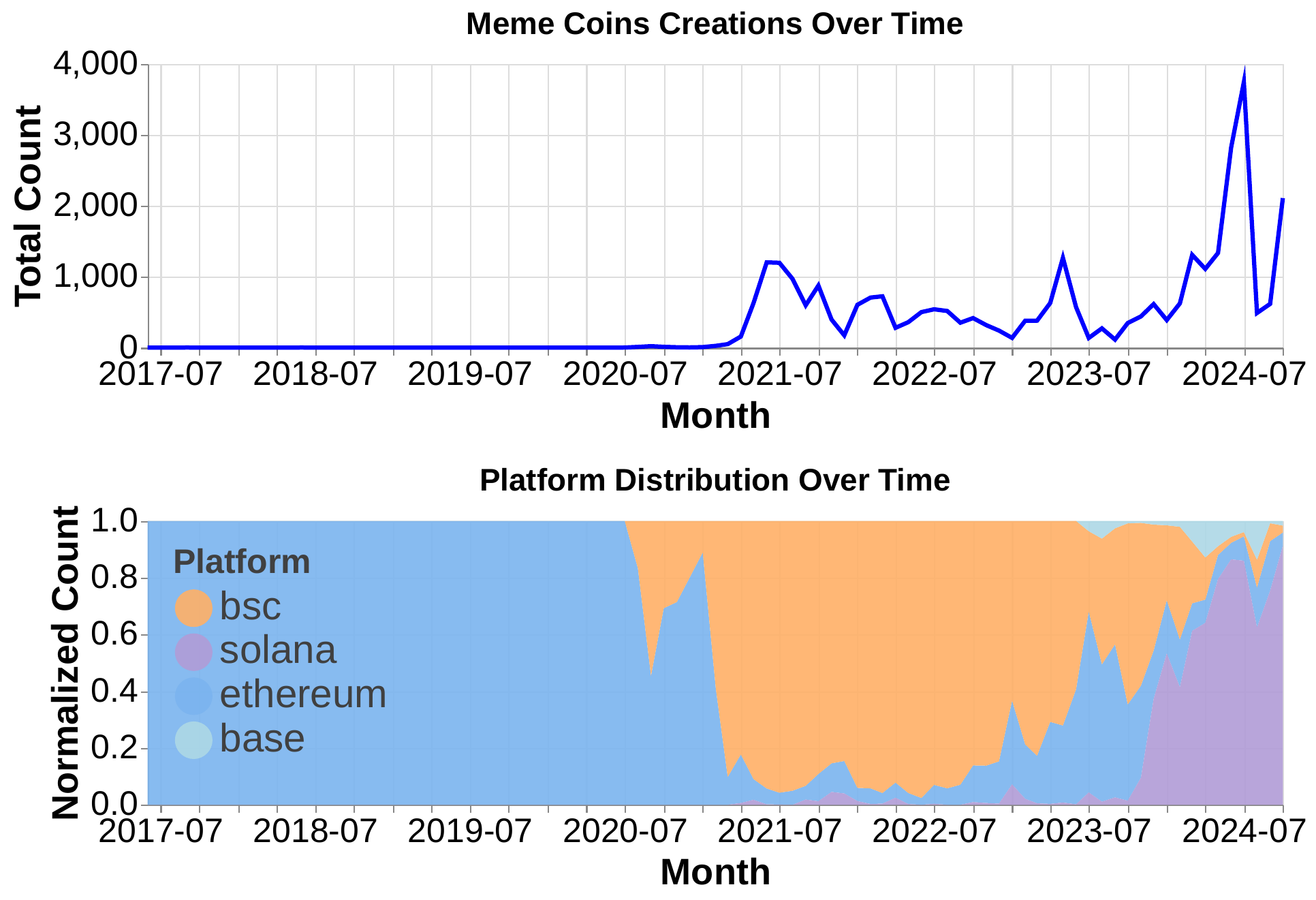}%
    }
    \caption{Creation date of meme coins over time and their distribution on different blockchains.}
    \label{fig:creation_date_analysis}
\end{figure}

Fig.~\ref{fig:creation_date_analysis} illustrates the temporal distribution of the 34,988 meme coins in the dataset. While this distribution reflects the specific snapshot of tokens we collected (potentially underrepresenting projects from previous years that have since disappeared), it captures distinct shifts in platform preference, from Ethereum’s early dominance to the rise of BNB Smart Chain (BSC), and subsequently to Solana and Base.
Within our sample, the period from 2017–2020 exhibits minimal activity, with fewer than 10 new tokens created per month. However, the subsequent surges registered align closely with pivotal ecosystem events.
The increase in deployments in late 2020 coincides with the cryptocurrency bull market and the introduction of BSC, where lower transaction fees likely catalyzed the volume of new projects~\cite{bull_run_2021, bsc_success}. Similarly, the activity spikes in 2023 correspond with the launch of Pepe Coin (April) and the single-sided staking mechanics introduced by BONK (October)~\cite{bonk_staking_pool}.
Moreover, our snapshot captures a pronounced surge in mid-2024, particularly on Solana. This peak coincides with the widespread adoption of the \textit{pump.fun} platform, suggesting that its simplified launch mechanisms significantly lowered the technical barriers represented in our data~\cite{pump_success}. We also analyzed naming conventions in meme coins, observing a shift from canine-themed tokens to AI and feline-related narratives in 2024 (see Sec.~\ref{appendix_sec: names_trend} in the Appendix).

\subsubsection{Lifecycle Analysis}
An essential characteristic of meme coins is their period of market relevance. To quantify this phenomenon, for each meme coin, we collected all its transactions' temporal data until mid-January 2025, approximately three months after our initial data collection period ended in mid-October 2024 (Sec.~\ref{subsec:dataset_overview}).
For EVM-compatible chains, we used their respective explorer APIs (EtherScan~\cite{etherscan_api}, BscScan~\cite{bscscan_api}, and BaseScan~\cite{base_scan}), whereas for Solana, we used the Alchemy APIs~\cite{alchemy_api}. From our original dataset of 31,811 tokens, we excluded 14 meme coins that had no recorded transactions, focusing our analysis on the remaining 31,797 tokens.

\mypara{Activity Patterns and Token Extinction.}
Our examination revealed that 15,099 meme coins (47.46\% of our sample) had no transaction activity after December 15, 2024, indicating they became inactive within just two months of our data collection period. The distribution of these abandoned tokens varied dramatically across blockchains (Tab.~\ref{tab:inactive_coins}): BSC exhibited the highest extinction rate, with 80.96\% of its meme coins (12,503 out of 15,442) showing no recent activity, followed by Ethereum with 37.32\% inactive tokens (1,388 out of 3,719) and Base with 16.20\% (134 out of 827). Solana demonstrated the highest resilience with only 9.09\% inactive tokens (1,074 out of 11,809).
It is worth noting, as shown in Fig.~\ref{fig:creation_date_analysis}, that the age distribution varies significantly across blockchains. The majority of tokens on Base and Solana were created in 2024, representing relatively new projects, while those on Ethereum and BSC have longer histories, providing a broader window for potential abandonment.

\mypara{The One-Day Meme Coin Phenomenon.}
Analyzing the lifetime of meme coins, defined as the interval between first and last recorded transactions, revealed alarming patterns of extremely short-lived tokens. Although our observation window limits the determination of the complete lifespan of recently created tokens, since we can only track them for three months, we identified 4,939 tokens (15.53\%) with a lifetime of 30 days or less and 3,512 tokens (11.04\%) with activity lasting only 7 days or less.
Most striking is the identification of 1,801 tokens (5.66\%) that ceased all trading activity within 24 hours of launch. We define these assets as One-Day Meme Coins. Among these ephemeral tokens, 276 recorded only a single transaction, while others showed varying levels of brief activity: 50\% had 7 or more transactions, and 10\% recorded more than 68 transactions before becoming inactive.
\begin{table}[ht]
\small
\centering
\caption{Distribution of inactive and one-day meme coins across blockchains.}
\label{tab:inactive_coins}
\begin{tabular}{lrrr}
\toprule
\textbf{Blockchain} & \textbf{Coins} & \textbf{Inactive Coins} & \textbf{1-Day Meme Coins}  \\
\midrule
BSC & 15,455 & 12,503 (80.96\%) & 1,598 (10.35\%)\\
Ethereum & 11,809 & 1,388 (37.32\%) & 129 (3.47\%)\\
Base & 827 & 134 (16.20\%) & 63 (7.62\%)\\
Solana & 3,720 & 1,074 (9.09\%) & 11 (0.09\%)\\
\midrule
\textbf{Overall} & 31,811 & 15,099 (47.46\%) & 1,801 (5.66\%)\\
\bottomrule
\end{tabular}
\end{table}
The prevalence of these one-day tokens is not uniformly distributed across blockchains. As shown in Table~\ref{tab:inactive_coins}, BSC dominated this category with 1,598 tokens (10.35\% of its total). This is followed by Base (7.62\%), Ethereum (3.47\%), and Solana with the lowest rate (0.09\%). These patterns suggest that certain blockchain ecosystems, particularly BSC, foster environments where tokens can rapidly lose market interest or become victims of rug pull schemes, as documented by~\cite{cernera2023token}.

\subsection{Digital Presence and Project Commitment}
\label{subsec:web_presence}
Web presence is a critical component for meme coin projects, serving both as a primary information source for potential investors and as a mandatory requirement for listing on influential cryptocurrency aggregators such as CoinMarketCap~\cite{coinmarketcap_listing} and CoinGecko~\cite{coingecko_listing}. Our analysis revealed that 23,755 out of 31,811 meme tokens in our dataset (74.80\%) claimed to have an associated website.

\subsubsection{Data Cleaning and URL Normalization}
To assess the authenticity of these digital presences, we first resolved all shortened URLs, resulting in 410 redirections across 303 distinct domains. %The most frequently used shortening services included odoo.com (32 occurrences), bit.ly (20), and t.co (14). 
We also identified 24 instances where the provided website was actually a Linktree page, a service that aggregates social media links. Manual examination revealed that those 24 Linktree pages, among others, contained links to legitimate project websites. 
Then, analyzing the 23,755 web pages, we identified 19,481 unique domains. Manual evaluation of domains with more than five occurrences (63 domains) revealed that many URLs did not lead to dedicated project websites but instead redirected to social media platforms or cryptocurrency services, as categorized in Tab.~\ref{tab:domain-types}.

\begin{table}[h]
\small
\centering
\caption{Domain Categories in Meme Coin Websites}
\label{tab:domain-types}
\begin{tabular}{lcc}
\toprule
\textbf{Domain Category} & \textbf{Count} & \textbf{Notable Examples} \\
\midrule
Social Media & 1,761 & Telegram (1,334), X (225) \\
Cryptocurr. Services & 807 & pump.fun (280), poocoin.app (178) \\
Other Pages & 74 & Wikipedia (11), KnowYourMeme (6) \\
\midrule
\textbf{Not a website} & 2,642 & \\
\midrule
Website Builders & 654 & vercel.app (79), canva.site (73) \\
Meme Related & 136 & claudeopusai (15), SHIBX.com (10) \\
\bottomrule
\end{tabular}
\end{table}

Our investigation uncovered 1,761 meme coins that redirected to social media platforms such as Telegram or X, and 807 tokens linked to cryptocurrency service domains that facilitate token creation (\eg pump.fun) or provide DeFi analytics (\eg poocoin.app). Additionally, 74 meme coins are linked to domains entirely unrelated to their projects and beyond their control, such as Wikipedia pages or official political websites.
Another category includes 654 websites built with website builders like vercel.app and canva.site, as well as meme coin-specific platforms like toekn.com and webby.fun/lobby.
Furthermore, 136 tokens use domains associated with other meme coins (\eg Doge-X.com or SHIBX.com, which are now for sale). An interesting case is \textit{claudeopusai.com}, which appears 15 times across different meme coins with varying names and original URLs, all ultimately redirecting to the website claudeopusai.com, the official website of another meme coin (\$OPUS). Similarly, the domain bonefloki.net appears 8 times across different meme coins despite their different names. %In contrast, domains such as anyinu.xyz and shiryoinu.online represent meme coins that share both name and website on the same blockchain.
After excluding the non-dedicated websites (2,642 URLs), our refined dataset contained 21,113 potential meme coin websites for deeper analysis.

\subsubsection{Infrastructure Investment and Persistence}
To evaluate the long-term commitment of meme coin projects, we examined domain registration metadata, web accessibility, and security aspects.

\mypara{Cost-Minimizing Registration Strategies.}
We successfully obtained WHOIS information for 11,819 websites (55.98\% of our refined dataset). The distribution of domain registrars revealed patterns consistent with cost-minimization strategies. The most common registrars were NAMECHEAP INC (3,251), GoDaddy.com, LLC (2,354), and Hostinger Operations, UAB (1,287), all known for their affordability. Another popular registrar was Porkbun (186 websites), known for its extremely low first-year pricing (often as low as \$1).
The preference for budget-oriented registrars suggests meme coin creators typically minimize infrastructure investments, consistent with short-term project visions. %Interestingly, we also identified five meme coins registered through the Cybersecurity and Infrastructure Security Agency, the U.S. registrar for government agencies. Further investigation revealed these projects had meme names such as \textit{NASA Musk}, \textit{PSYOP}, and \textit{CIA AI Bird Spy}.

\mypara{Ephemeral Web Existence.}
Our web accessibility testing revealed alarming rates of abandonment. Of the 21,113 websites, only 10,722 (50.78\%) were reachable via basic socket connection. Among those with successful connections, merely 6,766 (32.09\% of all purported websites) returned proper HTTP 200 status codes indicating functioning websites. 
This significant inaccessibility rate (67.91\%) provides compelling evidence that most meme coin projects lack long-term viability. The ephemeral nature of these websites aligns with the typical lifecycle of speculative tokens: Once initial excitement diminishes or developers achieve their financial objectives, projects are abandoned along with their digital infrastructure.

For the accessible domains, we successfully downloaded and archived the raw HTML source code to facilitate content-based forensic analysis. In particular, one specific source code (associated with the BaseAI token's web page) was flagged by a major commercial antivirus software as containing a potential JavaScript Trojan, highlighting the immediate security risks of the ecosystem.

\mypara{Security and Scam Analysis.}
To identify potentially malicious websites, we utilized Google Safe Browsing~\cite{safebrowsing} and Chain Patrol~\cite{chainpatrol}, specialized services that detect harmful sites and cryptocurrency scams. Google Safe Browsing flagged 9 websites as unsafe due to social engineering threats, while ChainPatrol blocked 8 web pages as phishing attempts.
Further investigation revealed limitations in these security tools. For instance, ChainPatrol flagged the website for \textit{Bitcoin Dogs} as a scam, but failed to detect \textit{Memeinator}, another documented scam~\cite{memeinator_scam_reddit} created by the same team. 

\subsubsection{Social Media Presence}
\begin{figure}
    \centerline{\includegraphics[width=0.46\textwidth]{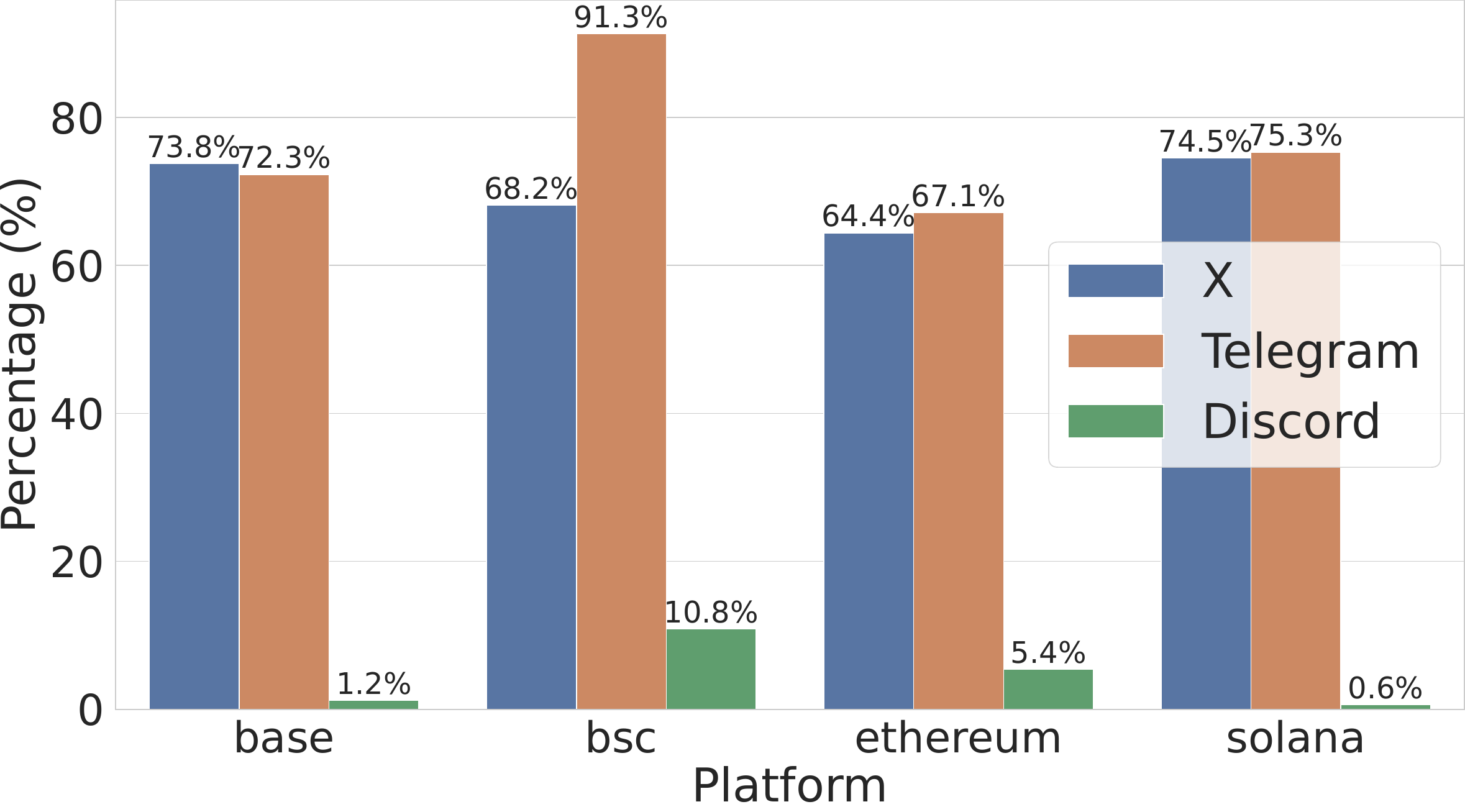}}
    \caption{Percentages of projects with social media presence.}
    \Description{A plot showing the percentages of meme coin projects having a social media account on X, Telegram, and Discord across Base, BSC, Ethereum, and Solana.}
  \label{fig:social_media_presences_ratio}
\end{figure}

Community engagement represents a critical aspect of meme coin ecosystems, with developers strategically leveraging online communication channels to attract investors~\cite{nani2022doge}.
The platforms referenced in Sec.~\ref{subsec:dataset_overview} allow developers to publish links to three social media platforms (Telegram, X, and Discord), as well as their project website, emphasizing the role of these communication channels in meme coin promotion.
To better understand these trends, we analyze the distribution of social media presence in our dataset. As shown in Fig.~\ref{fig:social_media_presences_ratio}, Telegram emerged as the dominant platform, utilized by 28,291 meme coin projects (80.85\% of the total), followed by X with 24,029 projects (68.67\%). Discord demonstrated significantly lower adoption, with only 2,344 projects (6.69\%) maintaining an active presence on this platform.

Telegram adoption was particularly pronounced on BSC, reaching 91.8\% of all BSC meme coin projects. Both Telegram and X maintained consistently high utilization rates across all blockchains, with adoption percentages exceeding 65\% in all cases. Instead, Discord usage remained minimal throughout the ecosystem, with especially low adoption on Solana (0.7\%) and Base (1.2\%). %Interestingly, BSC-based projects demonstrated substantially higher Discord utilization (11.4\%) compared to other blockchains, suggesting potentially different community building approaches within this ecosystem.

\section{Accessing the Dataset and FAIR Principles}
\label{sec:fair_principles}

We released \datasetname in strict alignment with the FAIR (Findable, Accessible, Interoperable, Reusable) principles~\cite{wilkinson2016fair} to ensure its longevity and utility for the research community.

\textbf{Findable.} 
The complete dataset~\cite{dataset_repository} is publicly hosted on Zenodo~\cite{zenodo}, a CERN-managed open repository that assigns a persistent Digital Object Identifier (DOI) to ensure citation stability. %Additionally, the associated collection scripts and documentation are maintained on GitHub~\cite{github_repo} to facilitate version control and community contributions.

\textbf{Accessible.} To maximize accessibility, the repository is organized into four logical modules. The Core and Financial Metadata modules provide a verified registry of contract specifications, validated timestamps, and market snapshots. These are complemented by the Digital Presence module, which catalogs forensic web data (WHOIS) and social endpoints (\eg Telegram, X), alongside the Raw Multimodal Assets archive, which stores the corresponding unstructured HTML source code and token logos.

\textbf{Interoperable.} 
All tabular data is provided in standard Comma-Separated Values (CSV) format. To facilitate efficient data transfer and handling, the raw unstructured assets (HTML source code and images) are aggregated into standard ZIP archives. Also, we follow a unified schema across all modules: every data point is indexed by a unique tuple, \texttt{(Contract Address, Blockchain Platform)}, allowing researchers to easily join financial, social, and visual data.

\textbf{Reusable.} 
The dataset is released under the Creative Commons Attribution 4.0 International (CC-BY 4.0) license, allowing for broad academic and commercial reuse provided proper credit is given. %We provide extensive documentation, including a detailed \texttt{README}, which outlines the data collection lineage, variable definitions, and the provenance of third-party security labels (e.g., ChainPatrol), ensuring that future users can accurately interpret the risk signals embedded in the data.

\section{Use Cases}
\label{sec:use_cases}
The multi-chain nature and rich metadata of \datasetname open several new avenues for research in data mining, cybersecurity, and financial forensics. %We highlight four primary applications where this dataset can serve as a critical resource.

\mypara{Multimodal Scam Detection.}
The utility of \datasetname is exemplified by recent research~\cite{mongardini2025midsummer}, which leveraged \datasetname longitudinal data to reveal that meme coin scams are not isolated events but sequential processes. The authors demonstrated how artificial growth strategies serve as necessary precursors to eventual exit scams.
However, detecting these patterns in real-time requires looking beyond the blockchain. \datasetname enables the development of new multimodal learning systems by providing the raw HTML source code and visual assets of project websites.
This data is critical for identifying \textit{phishing kits}, reusable, pre-packaged website templates that scammers deploy across different domains to clone legitimate projects.
Furthermore, the dataset supports \textit{visual similarity analysis} (\eg using Siamese networks) to detect brand impersonation, where tokens illicitly appropriate logos from established entities. Furthermore, the dataset facilitates \textit{incongruence detection}, allowing models to flag contradictions between a token's on-chain metrics and the "scarcity guaranteed" claims found on its landing page, a hallmark of fraudulent intent.

\mypara{Early Survival Prediction.}
A critical finding of our analysis is the prevalence of One-Day Meme Coins, meme coins that lose all transactional activity within 24 hours of deployment. These ephemeral assets frequently mask 1-day rug pulls, aggressive exit scams where developers drain liquidity immediately after launch, leaving investors with no time to react and resulting in total financial loss~\cite{cernera2023token}. 
Predicting this rapid mortality is therefore a high-stakes \textit{Survival Analysis} challenge. \datasetname provides granular timestamps (deployment time vs. last transaction) to train early-warning systems that assess token longevity. Crucially, the absence of a functional website often serves as an early warning sign: Many projects launch without any dedicated website, and among those that do, a significant fraction are currently unreachable, indicating rapid abandonment. By treating One-Day Meme Coins as a positive class for immediate risk, researchers can build classifiers that combine these web stability signals with transaction activity and social media presence to predict impending abandonment. Such models could serve as the basis for real-time investor protection tools that warn users about high-risk assets before they interact with the smart contract.

\mypara{Cross-Chain Study.}
Most decentralized finance (DeFi) research is currently siloed within the Ethereum ecosystem. By spanning Ethereum, BSC, Base, and Solana, \datasetname enables cross-chain graph-based studies of capital migration and fraud mechanisms. Researchers can leverage this dataset to analyze whether market manipulation patterns observed on Ethereum subsequently migrate to low-fee chains. Furthermore, the dataset supports the investigation of temporal correlations in meme trends. Researchers can identify the blockchain origin of a new popular naming trend, map its propagation across other ecosystems, and discover whether such trends consistently emerge from specific platforms. Understanding these cross-chain dynamics is fundamental for developing robust predictive market models that account for the interconnected nature of the meme coin ecosystem.

\mypara{Social Sentiment and Community Dynamics.}
Beyond structural links, \datasetname serves as a critical entry point for Natural Language Processing and Computational Social Science studies. By providing verified social media endpoints (Telegram, Discord, X), the dataset acts as a seed for collecting rich off-chain textual data. 
Researchers can utilize these handles to scrape community message logs, enabling the analysis of predatory marketing strategies. Specifically, this allows for the study of the tone and sentiment used by project administrators to lure investors, such as FOMO (Fear Of Missing Out) language, high-pressure urgency, or safety guarantees. 
Furthermore, the dataset enables correlating financial events with community metrics, such as subscriber counts and engagement velocity, allowing researchers to distinguish between organic community growth and artificial bot inflation.

\section{Ethical Considerations}
All data collected in this work were obtained from publicly accessible sources, including blockchain explorers and cryptocurrency aggregators. The dataset comprises only public blockchain records with no personally identifiable information.
Regarding data collection, we implemented rate limiting across all scrapers with a minimum interval of 5 seconds between requests to prevent service disruptions.
Moreover, we did not trade, create, or promote any tokens examined in this study, ensuring research independence and avoiding market manipulation. Finally, to preserve privacy, we deliberately avoided linking wallet addresses to real-world identities. While these addresses are public, we made no attempt to correlate them with external datasets (KYC databases) that could compromise user privacy.
Consequently, according to our IRB's policy, we did not need explicit authorization for our study.

\section{Limitations}
%\mypara{Temporal Scope.} 
Our dataset captures the meme coin ecosystem during mid-October 2024, with lifecycle data extending through January 2025. While this window enables longitudinal analysis of token mortality and One-Day Meme Coins, it may not reflect seasonal variations in meme coin creation patterns or capture longer-term survival dynamics beyond three months.
%\mypara{Data Availability Trade-offs.}
Moreover, to enable cross-chain analysis spanning four blockchains, we leveraged aggregator APIs. This approach sacrifices granular transaction-level data for breadth and computational feasibility. Researchers requiring complete on-chain traces should supplement our dataset with direct blockchain queries.
%\mypara{Classification Boundaries.}
Concerning the classification of meme coins, the name-based and pump.fun classification methods prioritize precision over recall to minimize the inclusion of non-meme tokens. Tokens with unconventional names that do not match our keywords may be underrepresented, particularly niche or emerging meme categories.
%\mypara{Website Snapshot Nature.}
Finally, the HTML archives of the web pages represent single-time snapshots collected in October 2024. Scam websites frequently evolve or disappear; thus, our dataset cannot capture dynamic changes in phishing tactics or post-collection modifications.

\section{Conclusions and Future Works}
\label{sec:conclusion}

This work introduces \datasetname, the first large-scale, cross-chain, and multi-modal repository designed to shed light on the meme coin ecosystem. By bridging the gap between on-chain mechanics and off-chain assets, we provided a granular view of the lifecycle of meme tokens.
Our analysis uncovered the prevalence of One-Day Meme Coins, ephemeral assets that experience no transactional activity after 24 hours of deployment. Correlated with this, we also observed the absence or rapid abandonment of project websites, which we identified as an early-stage indicator of low-effort scams and impending project failure.

%The release of \datasetname opens several avenues for future research. While current fraud detection relies heavily on transaction graphs, our dataset enables the development of multimodal classifiers that leverage on-chain metrics alongside semantic features from website HTML and visual embeddings from token logos. Furthermore, the raw HTML source code facilitates the detection of scam templates; researchers can apply code similarity analysis to cluster projects sharing identical web structures, potentially unmasking serial scam networks operating across different blockchains. Lastly, leveraging the granular timestamps in \datasetname, researchers can refine the survival analysis models proposed in this study. By treating One-Day Tokens as a distinct failure class, future models could estimate real-time failure probabilities for new projects, serving as a critical risk-assessment tool for investors navigating the decentralized landscape.

The release of \datasetname opens novel research directions. It supports multimodal classifiers that combine on-chain metrics with HTML semantics and visual embeddings. Additionally, the raw source code enables the detection of scam templates, allowing researchers to unmask serial fraud networks through code similarity. Finally, future work can refine survival models by treating One-Day Meme Coins as a distinct failure class to estimate real-time project risk.

\bibliographystyle{ACM-Reference-Format}
\bibliography{main}

@inproceedings{cernera2023token,
  title={Token Spammers, Rug Pulls, and Sniper Bots: An Analysis of the Ecosystem of Tokens in Ethereum and in the Binance Smart Chain (BNB))},
  author={Cernera, Federico and La Morgia, Massimo and Mei, Alessandro and Sassi, Francesco},
  booktitle={32nd USENIX Security Symposium (USENIX Security 23)},
  pages={3349--3366},
  year={2023}
}

@article{wilkinson2016fair,
  title={The FAIR Guiding Principles for scientific data management and stewardship},
  author={Wilkinson, Mark D and Dumontier, Michel and Aalbersberg, IJsbrand Jan and Appleton, Gabrielle and Axton, Myles and Baak, Arie and Blomberg, Niklas and Boiten, Jan-Willem and da Silva Santos, Luiz Bonino and Bourne, Philip E and others},
  journal={Scientific data},
  volume={3},
  number={1},
  pages={1--9},
  year={2016},
  publisher={Nature Publishing Group}
}

@inproceedings{wu2024tokenscout,
  title={Tokenscout: Early detection of ethereum scam tokens via temporal graph learning},
  author={Wu, Cong and Chen, Jing and Zhao, Ziming and He, Kun and Xu, Guowen and Wu, Yueming and Wang, Haijun and Li, Hongwei and Liu, Yang and Xiang, Yang},
  booktitle={Proceedings of the 2024 on ACM SIGSAC Conference on Computer and Communications Security},
  pages={956--970},
  year={2024}
}

@article{mongardini2025midsummer,
  title={A Midsummer Meme's Dream: Investigating Market Manipulations in the Meme Coin Ecosystem},
  author={Mongardini, Alberto Maria and Mei, Alessandro},
  journal={arXiv preprint arXiv:2507.01963},
  year={2025}
}

@book{shifman2013memes,
  title={Memes in digital culture},
  author={Shifman, Limor},
  year={2013},
  publisher={MIT press}
}

@inproceedings{ramos2003using,
  title={Using tf-idf to determine word relevance in document queries},
  author={Ramos, Juan and others},
  booktitle={Proceedings of the first instructional conference on machine learning},
  volume={242},
  number={1},
  pages={29--48},
  year={2003},
  organization={Citeseer}
}

@article{bholowalia2014ebk,
  title={EBK-means: A clustering technique based on elbow method and k-means in WSN},
  author={Bholowalia, Purnima and Kumar, Arvind},
  journal={International Journal of Computer Applications},
  volume={105},
  number={9},
  year={2014},
  publisher={Citeseer}
}

@article{kamps2018moon,
  title={To the moon: defining and detecting cryptocurrency pump-and-dumps},
  author={Kamps, Josh and Kleinberg, Bennett},
  journal={Crime Science},
  volume={7},
  number={1},
  pages={1--18},
  year={2018},
  publisher={Springer}
}

@article{stencel2023meme,
  title={What is a meme coin? Dogecoin to the moon!},
  author={Stencel, Adrian},
  journal={HAL science preprint: hal-04360574},
  year={2023}
}

@article{nani2022doge,
  title={The doge worth 88 billion dollars: A case study of Dogecoin},
  author={Nani, Albi},
  journal={Convergence},
  volume={28},
  number={6},
  pages={1719--1736},
  year={2022},
  publisher={SAGE Publications Sage UK: London, England}
}

@article{kim2024identifying,
  title={Identifying Networked Patterns in Memecoin Twitter Accounts Using Exponential Random Graph Modeling},
  author={Kim, Jae Hun and Park, Han Woo},
  journal={IT Professional},
  volume={25},
  number={6},
  pages={82--89},
  year={2024},
  publisher={IEEE}
}

@article{la2018pretending,
  title={Pretending to be a VIP! Characterization and Detection of Fake and Clone Channels on Telegram},
  author={La Morgia, Massimo and Mei, Alessandro and Mongardini, Alberto Maria and Wu, Jie},
  journal={ACM Transactions on the Web},
  volume={19},
  number={2},
  pages={1--24},
  year={2025},
  publisher={ACM New York, NY}
}

@inproceedings{la2025tgdataset,
  title={Tgdataset: Collecting and exploring the largest telegram channels dataset},
  author={La Morgia, Massimo and Mei, Alessandro and Mongardini, Alberto Maria},
  booktitle={Proceedings of the 31st ACM SIGKDD Conference on Knowledge Discovery and Data Mining},
  volume={1},
  pages={2325--2334},
  year={2025}
}

@inproceedings{la2023sa,
  title={It’sa Trap! detection and analysis of fake channels on telegram},
  author={La Morgia, Massimo and Mei, Alessandro and Mongardini, Alberto Maria and Wu, Jie},
  booktitle={2023 IEEE International Conference on Web Services (ICWS)},
  pages={97--104},
  year={2023},
  organization={IEEE}
}

@article{belcastro2023enhancing,
  title={Enhancing cryptocurrency price forecasting by integrating machine learning with social media and market data},
  author={Belcastro, Loris and Carbone, Domenico and Cosentino, Cristian and Marozzo, Fabrizio and Trunfio, Paolo},
  journal={Algorithms},
  volume={16},
  number={12},
  pages={542},
  year={2023},
  publisher={MDPI}
}

@article{long2024coinclip,
  title={CoinCLIP: A Multimodal Framework for Evaluating the Viability of Memecoins in the Web3 Ecosystem},
  author={Long, Hou-Wan and Li, Hongyang and Cai, Wei},
  journal={arXiv preprint arXiv:2412.07591},
  year={2024}
}

@article{long2024bridging,
  title={Bridging Culture and Finance: A Multimodal Analysis of Memecoins in the Web3 Ecosystem},
  author={Long, Hou-Wan and Wong, Nga-Man and Cai, Wei},
  journal={arXiv preprint arXiv:2412.04913},
  year={2024}
}

@article{gandal2018price,
  title={Price manipulation in the Bitcoin ecosystem},
  author={Gandal, Neil and Hamrick, JT and Moore, Tyler and Oberman, Tali},
  journal={Journal of Monetary Economics},
  volume={95},
  pages={86--96},
  year={2018},
  publisher={Elsevier}
}

@inproceedings{krafft2018experimental,
  title={An experimental study of cryptocurrency market dynamics},
  author={Krafft, Peter M and Della Penna, Nicol{\'a}s and Pentland, Alex Sandy},
  booktitle={Proceedings of the 2018 CHI conference on human factors in computing systems},
  pages={1--13},
  year={2018}
}

@inproceedings{daian2020flash,
  title={Flash boys 2.0: Frontrunning in decentralized exchanges, miner extractable value, and consensus instability},
  author={Daian, Philip and Goldfeder, Steven and Kell, Tyler and Li, Yunqi and Zhao, Xueyuan and Bentov, Iddo and Breidenbach, Lorenz and Juels, Ari},
  booktitle={2020 IEEE symposium on security and privacy (SP)},
  pages={910--927},
  year={2020},
  organization={IEEE}
}

@inproceedings{la2023game,
  title={A game of NFTs: Characterizing NFT wash trading in the Ethereum blockchain},
  author={La Morgia, Massimo and Mei, Alessandro and Mongardini, Alberto Maria and Nemmi, Eugenio Nerio},
  booktitle={2023 IEEE 43rd International Conference on Distributed Computing Systems (ICDCS)},
  pages={13--24},
  year={2023},
  organization={IEEE}
}

@article{cernera2024blockchain,
  title={The Blockchain Warfare: Investigating the Ecosystem of Sniper Bots on Ethereum and BNB Smart Chain},
  author={Cernera, Federico and La Morgia, Massimo and Mei, Alessandro and Mongardini, Alberto Maria and Sassi, Francesco},
  journal={ACM Transactions on Internet Technology (TOIT)},
  year={2024}
}

@inproceedings{zhou2024stop,
  title={Stop pulling my rug: Exposing rug pull risks in crypto token to investors},
  author={Zhou, Yuanhang and Sun, Jingxuan and Ma, Fuchen and Chen, Yuanliang and Yan, Zhen and Jiang, Yu},
  booktitle={Proceedings of the 46th International Conference on Software Engineering: Software Engineering in Practice},
  pages={228--239},
  year={2024}
}

@inproceedings{victor2021detecting,
  title={Detecting and quantifying wash trading on decentralized cryptocurrency exchanges},
  author={Victor, Friedhelm and Weintraud, Andrea Marie},
  booktitle={Proceedings of the Web Conference 2021},
  pages={23--32},
  year={2021}
}

@inproceedings{von2022nft,
  title={NFT wash trading: Quantifying suspicious behaviour in NFT markets},
  author={von Wachter, Victor and Jensen, Johannes Rude and Regner, Ferdinand and Ross, Omri},
  booktitle={International Conference on Financial Cryptography and Data Security},
  pages={299--311},
  year={2022},
  organization={Springer}
}

@inproceedings{hamrick2019economics,
  title={The economics of cryptocurrency pump and dump schemes},
  author={Hamrick, JT and Rouhi, Farhang and Mukherjee, Arghya and Feder, Amir and Gandal, Neil and Moore, Tyler and Vasek, Marie},
  booktitle={Workshop on the economics of information security},
  year={2019}
}

@article{la2023doge,
  title={The doge of wall street: Analysis and detection of pump and dump cryptocurrency manipulations},
  author={La Morgia, Massimo and Mei, Alessandro and Sassi, Francesco and Stefa, Julinda},
  journal={ACM Transactions on Internet Technology},
  volume={23},
  number={1},
  pages={1--28},
  year={2023},
  publisher={ACM New York, NY}
}

@inproceedings{la2020pump,
  title={Pump and dumps in the bitcoin era: Real time detection of cryptocurrency market manipulations},
  author={La Morgia, Massimo and Mei, Alessandro and Sassi, Francesco and Stefa, Julinda},
  booktitle={2020 29th international conference on computer communications and networks (ICCCN)},
  pages={1--9},
  year={2020},
  organization={IEEE}
}

@inproceedings{cernera2023ready,
  title={Ready, aim, snipe! analysis of Sniper Bots and their impact on the DeFi ecosystem},
  author={Cernera, Federico and La Morgia, Massimo and Mei, Alessandro and Mongardini, Alberto Maria and Sassi, Francesco},
  booktitle={Companion Proceedings of the ACM Web Conference 2023},
  pages={1093--1102},
  year={2023}
}

@inproceedings{xu2019anatomy,
  title={The anatomy of a cryptocurrency $\{$Pump-and-Dump$\}$ scheme},
  author={Xu, Jiahua and Livshits, Benjamin},
  booktitle={28th USENIX Security Symposium (USENIX Security 19)},
  pages={1609--1625},
  year={2019}
}

@misc{coingecko_listing,
  author = {CoinGecko},
  title = {Methodology},
  howpublished = {\url{https://www.coingecko.com/en/methodology}},
  year = {2025}
}

@misc{coinmarketcap_listing,
  author = {CoinMarketCap},
  title = {Listings Criteria},
  howpublished = {\url{https://support.coinmarketcap.com/hc/en-us/articles/360043659351-Listings-Criteria}},
  year = {2025}
}

@misc{chainpatrol,
  author = {ChainPatrol Inc. },
  title = {ChainPatrol},
  howpublished = {\url{https://chainpatrol.io/}},
  year = {2025}
}

@misc{safebrowsing,
  author = {Google},
  title = {Google Safe Browsing},
  howpublished = {\url{https://safebrowsing.google.com/}},
  year = {2025}
}

@misc{memeinator_scam_reddit,
  author = {r/CryptoScams},
  title = {Bitcoin Dogs \& Memeinator are both scams by the same group.},
  howpublished = {\url{https://www.reddit.com/r/CryptoScams/comments/1bqeos9/bitcoin_dogs_memeinator_are_both_scams_by_the/?rdt=64262}},
  year = {2024}
}

@misc{coinmakretcap,
  author = {CoinMarketCap},
  title = {Today's Cryptocurrency Prices by Market Cap},
  howpublished = {\url{https://coinmarketcap.com/}},
  year = {2025}
}

@misc{coinmakretcap_api,
  author = {CoinMarketCap},
  title = {CoinMarketCap API},
  howpublished = {\url{https://coinmarketcap.com/api/}},
  year = {2025}
}

@misc{coingecko,
  author = {CoinGecko},
  title = {Cryptocurrency Prices by Market Cap},
  howpublished = {\url{https://www.coingecko.com/}},
  year = {2025}
}

@misc{coingecko_api,
  author = {CoinGecko},
  title = {CoinGecko API},
  howpublished = {\url{https://www.coingecko.com/en/api}},
  year = {2025}
}

@misc{geckoterminal,
  author = {CoinGecko},
  title = {Decentralized Exchange (DEX) Tracker Tool},
  howpublished = {\url{https://www.geckoterminal.com/}},
  year = {2025}
}

@misc{dexscreener,
  author = {Dexscreener},
  title = {Dex Screener},
  howpublished = {\url{https://dexscreener.com/}},
  year = {2025}
}

@misc{coinsniper,
  author = {CoinSniper},
  title = {Best Coins Today},
  howpublished = {\url{https://coinsniper.net/}},
  year = {2025}
}

@misc{bnbsmartchain,
    author = {Bnbchain.org},
  title = {BNB Smart Chain - High Performance DeFi Hub},
  howpublished = {\url{https://docs.bnbchain.org/bnb-smart-chain/overview/}},
  year = {2024}
}

@misc{ethereum_doc,
    author = {Ethereum.org},
  title = {What is Ethereum?},
  howpublished = {\url{https://ethereum.org/en/learn/\#what-is-crypto-ethereum}},
  year = {2025}
}

@misc{base_doc,
    author = {Base.org},
  title = {Base Documentation},
  howpublished = {\url{https://docs.base.org/}},
  year = {2025}
}

@misc{solana_doc,
    author = {Solana Foundation},
  title = {Solana Documentation},
  howpublished = {\url{https://solana.com/docs}},
  year = {2025}
}

@misc{etherscan_api,
    author = {Etherscan},
  title = {Introduction},
  howpublished = {\url{https://docs.etherscan.io/}},
  year = {2025}
}

@misc{bscscan_api,
    author = {BscScan},
  title = {Introduction},
  howpublished = {\url{https://docs.bscscan.com/}},
  year = {2025}
}

@misc{base_scan,
    author = {BaseScan},
  title = {Introduction},
  howpublished = {\url{https://docs.basescan.org/}},
  year = {2025}
}

@misc{alchemy_api,
    author = {Alchemy Insights, Inc},
  title = {Solana API Quickstart},
  howpublished = {\url{https://docs.alchemy.com/reference/solana-api-quickstart}},
  year = {2025}
}

@misc{bull_run_2021,
    author = {Martin Young, Cointelegraph},
  title = {Top crypto winners and losers of 2021},
  howpublished = {\url{https://cointelegraph.com/news/top-crypto-winners-and-losers-of-2021}},
  year = {2021}
}

@misc{bonk_staking_pool,
    author = {Solana Compass},
  title = {What Is Bonk?},
  howpublished = {\url{https://solanacompass.com/projects/Bonk}},
  year = {2024}
}

@misc{pump_success,
    author = {Boaz Sobrado, Forbes},
  title = {From Memes To \$500 Million In Revenue: The Pump.Fun Phenomenon},
  howpublished = {\url{https://www.forbes.com/sites/boazsobrado/2025/02/22/from-memes-to-500-million-in-revenue-the-pumpfun-phenomenon/}},
  year = {2025}
}

@misc{bsc_success,
    author = {Coin98},
  title = {Binance Smart Chain 2021 Year In Review Report},
  howpublished = {\url{https://coin98.net/binance-smart-chain-2021-year-in-review}},
  year = {2023}
}

@misc{pump_fun_platform,
    author = {pump.fun},
  title = {Board},
  howpublished = {\url{https://pump.fun/board}},
  year = {2025}
}

@misc{dexscreener_listing,
  author = {DEX Screener},
  title = {Token Listing},
  howpublished = {\url{https://docs.dexscreener.com/token-listing}},
  year = {2024}
}

@misc{coinsniper_listing,
  author = {CoinSniper.Net},
  title = {New Coins},
  howpublished = {\url{https://coinsniper.net/new}},
  year = {2025}
}

@misc{solana_rpc,
  author = {Solana Foundation},
  title = {Clusters and Public RPC Endpoints},
  howpublished = {\url{https://solana.com/it/docs/references/clusters}},
  year = {2025}
}

@misc{meme_coins_sec,
  author = {U.S. Securities and Exchange Commission},
  title = {Staff Statement on Meme Coins},
  howpublished = {\url{https://www.sec.gov/newsroom/speeches-statements/staff-statement-meme-coins}},
  year = {2025}
}

@misc{meme_coins_definition_coingecko,
  author = {Coingecko},
  title = {The Global Crypto Classification Standard},
  howpublished = {\url{https://cdn.21shares.com/uploads/current-documents/reports/21Shares_R_TheGlobalCryptoClassificationStandard_September2024.pdf}},
  year = {2024}
}

@misc{meme_coins_definition_coinmarketcap,
  author = {CoinMarketCap},
  title = {Top Memes Tokens by Market Capitalization},
  howpublished = {\url{https://coinmarketcap.com/view/memes/}},
  year = {2025}
}

@misc{meme_success_cmc,
  author = {CoinMarketCap},
  title = {Meme coins are 2024’s most popular crypto narrative, capturing 31\% of global investor interest},
  howpublished = {\url{https://coinmarketcap.com/academy/article/fe803cfe-074d-4d3c-be34-91c0443b110a}},
  year = {2025}
}

@misc{impressive_return_meme_coins_spa,
  author = {Spa Coin},
  title = {10 Best Meme Coins to Buy Now: From DOGE to PEPE (2024 Guide)},
  howpublished = {\url{https://spacoin.io/best-meme-coins-to-buy/}},
  year = {2024}
}

@misc{dataset_repository,
  author = {Mongardini Alberto Maria and Mei Alessandro},
  title = {MemeChain: A Multimodal Cross-Chain Dataset for Meme Coin Forensics and Risk Analysis},
  howpublished = {\url{https://zenodo.org/records/18246856}},
  year = {2026}
}

@misc{zenodo,
  doi = {10.25495/7GXK-RD71},
  url = {https://www.zenodo.org/},
  author = {{European Organization For Nuclear Research} and {OpenAIRE}},
  language = {en},
  title = {Zenodo},
  publisher = {CERN},
  year = {2013}
}

\newpage

\appendix
\section{Names Trends Over Time.} 
\label{appendix_sec: names_trend}
\begin{figure*}
    \centerline{\includegraphics[width=0.86\textwidth]{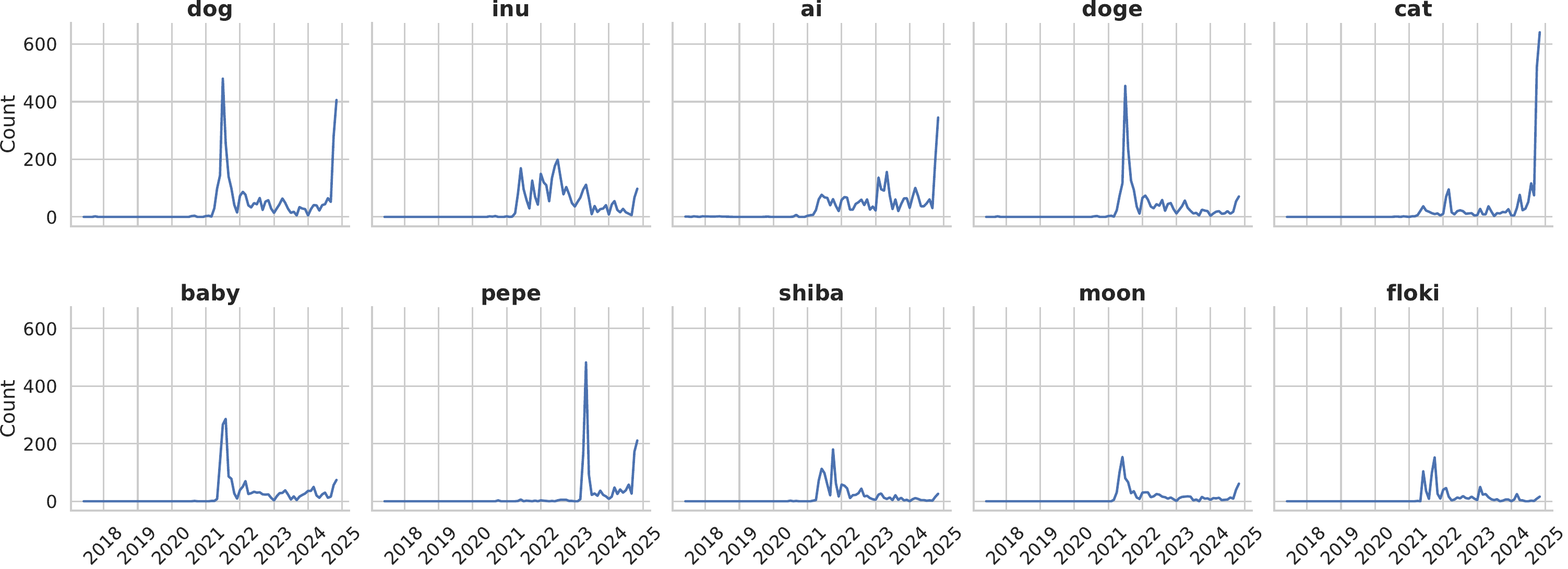}}
    \caption{Usage trends of the ten most popular words in meme coin names from 2018 to 2024 in our dataset.}
    \Description{A plot showing the usage trends of the ten most popular words in meme coin names from 2018 to 2024.}
  \label{fig:name_trend_over_time}
\end{figure*}

Fig.~\ref{fig:name_trend_over_time} shows the usage trends of the ten most popular words in meme coin names over time in our dataset, revealing distinct patterns in meme coin naming conventions between 2018 and 2024, reflecting cultural phenomena, market influences, and emerging technological themes.

Canine-related terminology represents a persistent pattern in meme coin naming conventions across the study period. The keyword \textit{dog} demonstrated significant popularity spikes in 2021 and 2024, establishing itself as a recurring theme in the meme coin ecosystem. Similarly, \textit{inu} (Japanese for dog) exhibited consistent growth post-2020, with periodic intensity increases. \textit{Doge} peaked notably in 2021, coinciding with Dogecoin's mainstream breakthrough, and maintained a steady presence thereafter. \textit{Shiba} also reached maximum popularity in 2021, aligning with the Shiba Inu token's rise, though its usage declined in subsequent years. \textit{Floki} saw peak adoption during 2021-2022, likely influenced by Elon Musk's public references to this name.

Another key aspect of meme coin names is their connection to emerging themes. 
\textit{AI} showed a dramatic usage increase in 2024, indicating a clear shift toward incorporating artificial intelligence narratives into meme coin branding. Concurrently, \textit{cat} experienced a sharp upward trajectory beginning in 2024, potentially signaling the emergence of feline-themed alternatives to the previously dominant canine motifs.

Also, cultural references significantly influenced naming trends. An example of this is \textit{Pepe} that underwent a sudden, pronounced spike in 2023, demonstrating the rapid adoption of this internet meme character in token naming. The term \textit{baby} peaked in 2021, representing a brief but significant naming trend. Finally, \textit{Moon}, commonly associated with ambitious price targets in cryptocurrency communities, peaked around 2021 before declining in subsequent periods.
These temporal naming patterns illustrate how meme coins' names are dynamically influenced by contemporary cultural events, market sentiment, and emerging technological paradigms.

\end{document}